\pdfoutput=1
\documentclass[
]{ceurart}

\usepackage{listings}
\usepackage{amsmath,amssymb}
\usepackage{enumitem}
\usepackage{graphicx}

\newcommand{\E}{\mathbb{E}}
\newcommand{\Prob}{\mathbb{P}}
\newcommand{\ind}{\mathbf{1}}

\newcommand{\pcur}{p_{\mathrm{cur}}}
\newcommand{\pR}{p_{R}}
\newcommand{\pFP}{p_{\mathrm{FP}}}
\newcommand{\pBIN}{p_{\mathrm{BIN}}}
\newcommand{\phat}{\hat{p}_{f}}

\usepackage{booktabs}

\begin{document}

\copyrightyear{2026}
\copyrightclause{Copyright for this paper by its authors.
  Use permitted under Creative Commons License Attribution 4.0
  International (CC BY 4.0).}

\conference{ECOM'26: SIGIR Workshop on eCommerce, Jul 24, 2026, Melbourne, Australia}

\title{Marginal Expected Revenue for Jointly Ranking Auction and Fixed-Price Listings in E-Commerce Sponsored Search}

\author[]{Greg Kocher}[%
  orcid=0009-0004-4832-1580,
  email=gkocher@ebay.com,
]
\cormark[1]
\author[]{Sanjana Arun}[%
  orcid=0009-0008-3710-3958,
  email=sanjanaarun@ebay.com,
]
\address[]{eBay, USA}

\cortext[1]{Corresponding author.}

\begin{abstract}
E-commerce search ranking must balance multiple objectives---relevance, user engagement, and platform revenue---when allocating impression slots to competing listings. Estimating the expected revenue component is well understood for fixed-price items, but becomes challenging when marketplace inventory includes mixed listing formats such as pure auctions and hybrid ``Auction with Buy It Now'' (ABIN) items, where prices evolve dynamically and the final transaction value is unknown at ranking time. Yet auction and ABIN listings account for a meaningful share of inventory and transaction volume on platforms such as eBay, and are a popular format for individual sellers and for unique items with unclear value.
We extend the standard Expected Cost-per-Mille (eCPM) framework to auction and ABIN listings by deriving a \textbf{marginal eCPM (meCPM)} that captures the incremental value of showing one more impression of an item whose price is still evolving. The resulting formulation extends the familiar fixed-price eCPM---which is already inherently marginal---to auction dynamics, allowing unified ranking of fixed-price, auction, and ABIN listings under a single objective. We then describe a practical production implementation that approximates this objective, addressing cold-start challenges by bootstrapping from existing engagement models.
Online A/B tests at a large e-commerce platform showed positive revenue gains and statistically significant improvements to user metrics, and the system was deployed to production.
\end{abstract}

\begin{keywords}
  e-commerce \sep
  sponsored search \sep
  ad ranking \sep
  auction listings \sep
  expected revenue \sep
  eCPM
\end{keywords}

\maketitle

\section{Introduction}
E-commerce search ranking systems must allocate limited impression slots across competing listings to maximize user and platform objectives such as buyer engagement and revenue. When a buyer searches for a product, the platform aims to provide the user with the most relevant and high quality results for their search query. Meanwhile, sellers want to sell their item quickly and at a favorable price. To this end, some sellers may enroll their item in an advertising program in order to boost their item to gain extra visibility, while paying the platform a fee which depends on the type of ad program and certain details about the listing. 
Two common ad program types in sponsored search are Cost-per-Click (CPC)~\cite{edelman2007gsp,varian2007position}, where the seller pays the platform a small fee each time a buyer clicks on the promoted item, and Cost-per-Acquisition (CPA)~\cite{mahdian2007ppa}, where the seller pays the platform a predetermined fraction of the final selling price (the ``ad rate'') only if the item sells.
Finally, the platform aims to facilitate and monetize each transaction.
This three-party interaction---buyer, seller, and platform---creates a multi-objective optimization problem where the search ranking system must balance relevance and user satisfaction against seller visibility and platform revenue, all within the tight latency and scale constraints of real-time search.
This is typically implemented through multiple stages, including retrieval~\cite{saha2025keyword}, followed by several rounds of filtering~\cite{hillard2010adrelevance,chaudhary2025quality}, re-ranking, blending~\cite{carrion2023virtual}, de-duplicating, and layout optimization, which condense a large set of candidate items into a final ranked list on the Search Results Page (SRP).

In sponsored search systems, part of this overall allocation process is often driven by value-aware ranking models that estimate expected revenue per impression, also referred to as Expected Cost-per-Mille (eCPM)~\cite{rosales2012pcc,yuan2019ecpm,zhao2023copr}. This scoring is well-understood in \emph{fixed-monetization} settings, which include both CPC (where the advertiser's bid per click is known in advance) and fixed-price CPA (where the item's selling price and ad rate are known), and where engagement---a click or a sale---is modeled as a binary outcome.
However, at some marketplaces like eBay, a notable portion of inventory operates under auction dynamics, where prices evolve over time and final transaction value is unknown at ranking time. Auction-based listings account for a meaningful share of inventory and serve an important role in the marketplace: they provide a price discovery mechanism for unique, one-of-a-kind, or hard-to-value items where the seller cannot easily set a fixed price~\cite{krishna2009auction,chen2017seller}, and are particularly popular among individual and consumer-to-consumer (C2C) sellers~\cite{einav2018auctions}. Yet these listings have historically been out of scope of CPA advertising programs due to the difficulty of estimating their value in a comparable way. For fixed-monetization programs---both CPC and fixed-price CPA---there is a single source of uncertainty (whether the buyer will engage), while the monetization value conditional on engagement is known in advance.
For CPA auction items, however, there are \emph{two} sources of uncertainty: not only whether a buyer will engage, but also what the final selling price (and therefore the platform's monetization) will be. This creates the need for new approaches to estimate and compare expected revenue across all listing formats, even when that term is only one ingredient of a broader multi-objective ranker. 

In summary, our main contributions are:
\begin{enumerate}[nosep]
    \item We extend the standard eCPM framework to auction listings by introducing a \textbf{marginal eCPM (meCPM)} that accounts for the stochastic, evolving price dynamics of proxy-bidding auctions.
    \item We derive meCPM expressions for fixed-price, pure auction, and hybrid Auction with Buy It Now (ABIN) listings, enabling a unified comparison of all listing formats on a revenue basis.
    \item We present a practical, simplified implementation that approximates these formulations and still leads to generally well-calibrated revenue prediction, allowing us to address cold-start challenges and launch a new auction CPA ad program.
    \item We validate the approach in online A/B tests and achieve statistically significant user metrics and positive revenue gains, and have deployed a version of the system to production.
\end{enumerate}

\section{Background and Related Work}

Sponsored search ranking in e-commerce platforms typically involves scoring each candidate ad item by its expected revenue per impression (eCPM), and then ranking items directly by revenue~\cite{aryafar2017ensemble,yang2025walmart,lu2014revenue};
using multi-objective scoring functions~\cite{carmel2020multiobj,kolbin2025unified,louca2019joint};
introducing explicit constraints on relevance, quality, inventory, or performance~\cite{malgireddy2025profit,chen2011rtb,luzhang2024lp};
or estimating the incremental (causal) value of showing each ad~\cite{xu2016lift,he2024uplift,bottou2013counterfactual}.
Formally, the expected revenue per impression has the general form~\cite{rosales2012pcc,yuan2019ecpm,chen2011rtb,lee2012cvr}:
\begin{equation}
\mathrm{eCPM} \;=\; \Prob(\text{action}\mid\text{impression}) \;\times\; R_{\text{action}}
\label{eq:ecpm_generic}
\end{equation}
where throughout the paper, we omit the conventional factor of 1000 used in Cost-per-Mille notation, and where ``action'' is the monetizing event and $R_{\text{action}}$ is the revenue earned by the platform when the action occurs.
For CPC ads, the action is a click and the revenue per click is an advertiser's bid.
For fixed-price CPA, the action is a sale and the revenue per sale is $p_{\mathrm{FP}} \cdot r$, where $p_{\mathrm{FP}}$ is the item's fixed price and the ad rate $r$ is the fraction of the final selling price the seller pays to the platform~\cite{ge2021relevance}.
Thus,
\begin{equation}
\mathrm{eCPM}_{\mathrm{CPC}} \;=\; \Prob(\text{click}\mid\text{impression}) \;\times\; \mathrm{bid}
\label{eq:ecpm_cpc}
\end{equation}
\begin{equation}
\mathrm{eCPM}_{\mathrm{FP}} \;=\; \Prob(\text{sale}\mid\text{impression}) \;\times\; p_{\mathrm{FP}} \;\times\; r
\label{equation:fp_ranking_score}
\end{equation}

In both of these cases, the outcome is binary---a click happens or not, a sale happens or not---and the revenue gained from the impression, $R_{\text{action}}$, is fixed and known in advance and is realized immediately when the action is taken.
Auction listings do not fit this pattern: the final price is unknown at ranking time, and although the transaction is not funded until auction close, a portion of the eventual revenue is already ``locked in'' by past bids and should not be attributed to future impressions. Therefore, a reasonable ranking objective is the \emph{marginal} expected revenue, or \emph{meCPM}: the incremental revenue that one more impression generates above the locked-in baseline.
This distinction conceptually mirrors lift-based approaches to ad ranking~\cite{xu2016lift,he2024uplift,bottou2013counterfactual} and the general economic definition of marginal revenue~\cite{mankiw2024principles}.
For CPC and fixed-price CPA, because revenue gain is ``all or none'' and there is no ``locked-in'' partial revenue accumulating over time, the standard definition of expected revenue per impression \emph{is} the \emph{marginal} expected revenue: the value of showing one more impression equals the full eCPM, so $\mathrm{meCPM}_{\mathrm{CPC}} = \mathrm{eCPM}_{\mathrm{CPC}}$ and $\mathrm{meCPM}_{\mathrm{FP}} = \mathrm{eCPM}_{\mathrm{FP}}$.
Similarly, pure Cost-per-Impression ads (CPI, or CPM if scaled by 1000) are a trivial special case, where the impression itself is the action, and by definition, $\mathrm{meCPM}_{\mathrm{CPI}}=\mathrm{eCPM}_{\mathrm{CPI}}$.
In the following section, we derive the marginal eCPM for CPA auction and ABIN listings under proxy bidding mechanics (such as at eBay), and show that expected revenue scoring of all listing formats fits into the marginal eCPM framework.

\section{Methods}
\subsection{Pure Auction Format}
\label{sec:pure_auction}
\paragraph{Proxy Bidding Mechanism}
In \emph{proxy bidding} (automatic bidding) systems, as used in eBay auctions, when a bidder places a bid, they submit the maximum value they are willing to pay, $V$, not a visible bid amount. The system automatically bids on their behalf, and on behalf of all previous bidders, raising the current visible price, $\pcur$, only as high as necessary to give the bidder with the highest valuation the leading position in the auction. How much $\pcur$ increases depends on the values of several parameters (see (\ref{eq:delta_pcur_piecewise})): the new bid amount, $V$; the current leader's hidden bid amount, $M$; and the minimum bid increment, $\delta$, which determines how much higher $V$ must be relative to the current price $\pcur$ (see Appendix~\ref{app:notation}).
Bids accumulate during the lifetime of the auction (typically 3, 7, or 10 days), and at auction close, the winner is the bidder with the highest submitted maximum, and they pay the second-highest submitted maximum (plus a small amount, which depends on $V$, $M$, and $\delta$).
This is similar in outcome to a second-price sealed-bid (Vickrey) auction~\cite{krishna2009auction,vickrey1961}, and in this scenario, bidders are incentivized to submit their true maximum willingness to pay, and there is no incentive for strategic \emph{bid shading}~\cite{rileysamuelson1981}. However, our derivation does not require truthfulness. We only require that $V$ be the submitted maximum bid used by the proxy-bidding mechanism. This distinction is important because in practice, the visible ascending price and fixed close-time give rise to \emph{auction sniping} behavior, where bids are strategically withheld until immediately before the auction ends~\cite{roth2002snipe}. Although not our main focus, the framework can in principle capture sniping behavior via the time-dependence of auction state $s$, which conditions both the bid-through-rate $\Prob(\mathrm{valid\ bid}\mid\mathrm{imp},s)$ and the conditional CDF $F_s$ from which $V$ is drawn.

\paragraph{Locked-in Revenue}
In ascending auctions, the price increases as new bids arrive. This leads to a key fact: if the current visible price is $p_{\mathrm{cur}}$, then the platform has already \emph{locked-in} revenue of $p_{\mathrm{cur}} \times r$, which will be realized regardless of future impressions. An additional impression should therefore compete based on the \emph{incremental} revenue it is expected to generate above this locked-in amount.

\paragraph{Greedy Next-Impression Approximation}
We compute the marginal value of only the single next impression, yielding a greedy, closed-form expression that depends on $M$, $\delta$, the observable auction/context state $s$, and the conditional bid-value CDF $F_s$ of submitted maximum bids given that a valid bid occurs. A richer treatment could model the full remaining impression sequence until auction close, e.g.\ via dynamic programming with Bellman recursions over the auction state.

\subsubsection{Marginal eCPM Derivation}

Let $s$ denote the observable auction/context state, let $p_0$ denote the seller's starting bid price, and let $\tau$ denote the valid-bid threshold: $\tau = p_0$ before any bid has been placed, and $\tau = \pcur + \delta$ once the auction has an active bid.
The derivation in this subsection assumes the post-bid setting, in which $\tau = \pcur + \delta$. The bid-through-rate model estimates $\Prob(\mathrm{valid\ bid}\mid \mathrm{imp},s)$. For a valid bid ($V \geq \tau$), i.e., a bid high enough to be successfully submitted to the platform, the price movement is (Appendix \ref{app:pure_auction_derivation_price_change_cases}):
\begin{equation}
\Delta \pcur(V; \pcur, M, \delta)
\;=\;
\left\{
\begin{array}{lll}
(V + \delta) - \pcur & \text{if } \pcur + \delta \leq V \leq M - \delta & (\pcur \to V + \delta) \\[4pt]
M - \pcur & \text{if } M - \delta < V \leq M & (\pcur \to M) \\[4pt]
V - \pcur & \text{if } M < V < M + \delta & (\pcur \to V) \\[4pt]
(M + \delta) - \pcur & \text{if } V \geq M + \delta & (\pcur \to M + \delta)
\end{array}
\right.
\label{eq:delta_pcur_piecewise}
\end{equation}
Submitted maxima below $\tau$ would not be valid bids and would produce no price change. Conditional on a valid bid occurring, the bidder submits a maximum $V$ drawn from the distribution with conditional CDF $F_s$, where $F_s(v)=\Prob(V\le v\mid \mathrm{valid\ bid},s)$ --- i.e.\ $F_s$ is supported on valid submitted maxima given the current state, $s$.
Note that although $\delta$ is piecewise-constant in price, here we treat it as a single value at the current $\pcur$; this is exact within a tier and an approximation at tier boundaries.

Taking the expectation of \eqref{eq:delta_pcur_piecewise} over the bidder's maximum $V$, and multiplying by the ad rate $r$ and the probability that the impression produces a valid bid, gives the marginal eCPM in factored form:
\begin{equation}
\text{meCPM}_{\text{auction}}
\;=\;
r \cdot \Prob(\mathrm{valid\ bid}\mid\mathrm{imp},s) \cdot
\E_{V\sim F_s}\!\left[\Delta \pcur(V;\pcur,M,\delta)\right],
\label{eq:pure_auction_factored}
\end{equation}
where $\Prob(\mathrm{valid\ bid}\mid\mathrm{imp},s)$ is the \emph{bid-through-rate}---the
probability that a given impression results in a valid bid, conditional on features of the item, context, time, etc.---and the second factor is the expected price
uplift given that a valid bid is placed.
We expand the conditional expectation using \eqref{eq:delta_pcur_piecewise} (full derivation in
Appendix~\ref{app:pure_auction_derivation}), and define the conditional clipped survival integral $\mathcal{J}_s(x,y) \equiv \int_x^y S_s(v)\,dv$ when $x < y$ and $\mathcal{J}_s(x,y) = 0$ when $x \geq y$, where $S_s(v) = 1 - F_s(v)$ is the conditional survival function of submitted maximum bids given a valid bid in state $s$. In the post-bid setting (where $\tau = \pcur+\delta$, and $M$ is defined since there is an existing leader):
\begin{equation}
\begin{aligned}
\mathrm{meCPM}_{\mathrm{auction}}
\;=\; & r \cdot \Prob(\mathrm{valid\ bid}\mid\mathrm{imp},s) \cdot \Bigl[ \bigl(p_{\mathrm{new}}(\tau) - \pcur\bigr) \\
& \quad +\; \mathcal{J}_s(\tau,\,M{-}\delta) \;+\; \mathcal{J}_s\bigl(\max(\tau,M),\,M{+}\delta\bigr) \Bigr]
\end{aligned}
\label{eq:pure_auction_exact}
\end{equation}
where $p_{\mathrm{new}}(\tau)$ is the visible price after the minimum valid bid, given by $\min(M, \tau + \delta)$ when $\tau \leq M$ and $\min(\tau, M + \delta)$ when $\tau > M$ (see \eqref{eq:pnew}).
The three terms inside the bracket have intuitive interpretations.
\textbf{(i)}~$(p_{\mathrm{new}}(\tau) - \pcur)$ is the \emph{baseline uplift} from the minimum valid bid, conditional on a valid bid occurring.
\textbf{(ii)}~$\mathcal{J}_s(\tau, M{-}\delta)$ captures additional upside when the old leader with hidden maximum $M$ keeps the lead and the visible price is pushed up to $V + \delta$.
\textbf{(iii)}~$\mathcal{J}_s(\max(\tau,M), M{+}\delta)$ captures the case where the challenger barely takes the lead and the visible price moves just high enough to overtake $M$.

A few observations about this expression are worth noting (see Appendix~\ref{section:mecpm_auction_derivation_sanity_checks}).
Among the three bracket terms, the baseline uplift $(p_{\mathrm{new}}(\tau) - \pcur)$ and the third term are each $O(\delta)$, while the second term $\mathcal{J}_s(\tau, M{-}\delta)$ scales with $M - \pcur$ and can dominate when $M$ is significantly above $\pcur$, reflecting the substantial expected upside from items with hidden headroom for the price to climb.
Conditional on a valid bid, a higher current visible price $\pcur$ reduces the expected price uplift from the next bid (the bracket term of \eqref{eq:pure_auction_exact}), although the full meCPM also depends on how $\Prob(\mathrm{valid\ bid}\mid\mathrm{imp},s)$ and the conditional bid-value distribution $F_s$ vary with the auction state. In practice, items whose visible price has already risen close to their typical final price have lower marginal value per impression.
Holding the valid-bid distribution and bid-through-rate fixed, the price-uplift expression is increasing in $M$: a larger hidden maximum from the current leader gives more room for the visible price to climb.

The zero-bid starting case (no bids yet, $\pcur = p_0$, no locked-in revenue) is simply $\mathrm{meCPM}_{\text{zero-bid}} = r \cdot \Prob(\mathrm{valid\ bid}\mid\mathrm{imp},s) \cdot p_0$, i.e.\ the full settlement price, since any valid first bid becomes the leader at visible price $p_0$. See Appendix~\ref{app:zero_bid}.

\subsubsection{Reserve Prices}
\label{sec:reserve_prices}

Some auctions have a hidden reserve price $p_R$, a seller-specified threshold below which the item will not sell even if the auction closes with a winning bid. Bidders see whether the reserve has been met but not its value~\cite{rileysamuelson1981,bajari2003winners}.
The reserve modifies the meCPM formulation. When at least one valid bid has been placed and the reserve has been met (so $\pcur \geq \pR$), the item will sell, locked-in revenue is $\pcur \times r$ as in the no-reserve case, and meCPM reduces to the post-bid pure-auction case in \eqref{eq:pure_auction_exact}.
If $\pcur < \pR$, the item is not yet guaranteed to sell, so locked-in revenue is \emph{zero}, not $\pcur \times r$. Unlike the no-reserve derivation above where the zero-bid case needed a separate formula, here both the post-bid and zero-bid starting states are addressed in a single formula.
A reserve-clearing bid ($V \geq \pR$) gives a settlement price of $\min(V, C)$, where the settlement cap $C$ is piecewise: $C \equiv \max\{\pR, M{+}\delta\}$ post-bid, and $C \equiv \pR$ zero-bid (when $M$ is undefined), and only valid bids with $V \geq L$, where $L \equiv \max\{\pR, \tau\}$, also clear the reserve. Taking the expectation over such bids (full derivation in Appendix~\ref{app:reserve_derivation}) we get:
\begin{equation}
\text{meCPM}_{\text{reserve not met}}
\;=\;
r \cdot \Prob(\mathrm{valid\ bid}\mid\mathrm{imp},s) \cdot
\left[\min(L,C)\,S_s(L) \;+\; \mathcal{J}_s(L,\,C)\right]
\label{eq:reserve_not_met}
\end{equation}
where $L \equiv \max\{\pR,\tau\}$ and $C$ is given by the piecewise definition above, and $S_s(L) \equiv \Prob(V>L\mid \mathrm{valid\ bid},s)$ is the conditional survival probability that a valid bid also clears the reserve.
In the common case $\pR \geq M + \delta$ (and trivially in the zero-bid case via the piecewise definition of $C$), we have $C = \pR \leq L$, so $\mathcal{J}_s(L,C) = 0$ and the formula reduces to $r \cdot \Prob(\mathrm{valid\ bid}\mid\mathrm{imp},s) \cdot S_s(L) \cdot \pR$, i.e.\ the reserve settlement price (scaled by the ad rate) weighted by the probability that the impression produces a reserve-clearing bid.

\subsection{ABIN Format}

An Auction with Buy It Now (ABIN) listing combines auction and fixed-price mechanics. The item runs as an auction with a starting bid but also has a Buy It Now price $p_{\mathrm{BIN}}$ at which a buyer may purchase immediately, ending the auction~\cite{wang2008bin}. So, a single impression can trigger at most one of two mutually exclusive revenue generating events: a BIN purchase or a new bid. The marginal eCPM becomes:
\begin{equation}
\mathrm{meCPM}_{\mathrm{ABIN}} \;=\; r \cdot \Prob(\text{BIN}\mid\text{imp},s) \cdot \pBIN \;+\; \mathrm{meCPM}_{\mathrm{auction}}(s)
\label{eq:abin_marginal}
\end{equation}
where $\mathrm{meCPM}_{\mathrm{auction}}(s)$ is the auction-format meCPM appropriate to the listing's current state: the reserve-not-met case \eqref{eq:reserve_not_met} when a reserve is set and unmet;
the post-bid pure-auction case \eqref{eq:pure_auction_exact} when at least one valid bid exists and any reserve is met; or the zero-bid no-reserve form $r \cdot \Prob(\mathrm{valid\ bid}\mid\mathrm{imp},s) \cdot p_0$ when no bid has been placed and no reserve applies (Appendix~\ref{app:zero_bid}).
On eBay, the BIN option typically disappears once the first bid is placed, so $\Prob(\text{BIN}\mid\text{imp},s) = 0$ post-bid and the ABIN formula collapses to the auction case.
We note the corner case where for some categories the BIN option remains after getting bids, until $\pcur$ reaches a fraction of $\pBIN$.
In that case, if there is no reserve price, or if the reserve is already met, then $\pcur \cdot r$ is already locked in so the first term of \eqref{eq:abin_marginal} instead uses $(\pBIN-\pcur)\cdot r$.

\subsection{Unified Ranking Across Listing Formats}

All listing formats share the meCPM structure $r \times \Prob(\text{event}\mid\text{imp}) \times \text{price signal}$. For CPI, CPC, and fixed-price CPA, there is no locked-in partial revenue, so the standard eCPM is already marginal ($\mathrm{meCPM} = \mathrm{eCPM}$). The full meCPM equations (\ref{eq:pure_auction_exact}), (\ref{eq:reserve_not_met}), and (\ref{eq:abin_marginal}), extend this to incrementally evolving prices, and in principle, along with the standard CPC~\eqref{eq:ecpm_cpc} 
and fixed-price CPA~\eqref{equation:fp_ranking_score} equations, could be used to score all listing formats on a revenue basis and fit them into the overall ranking system with other terms and constraints.
However, this requires multiple components not available at launch time due to the cold-start setting.

\subsection{Simplified Implementation}
\label{sec:simplification}

We explain several practical simplifications to enable a deployable system which brings statistically significant gains (Section~\ref{sec:online_results}),
discuss alternative modeling approaches, and leave the full meCPM implementation as future work.
First, although a proxy-bidding platform may in principle have access to $M$ or $\pR$ if present, a ranking implementation may prefer not to require them, since a score which depends on $M$ or $p_R$ could, in theory, subtly reveal information about these values.
Alternative treatments of $M$ and $\pR$ are left for future work, and for now we omit these quantities.
Second, $\delta$: in many e-commerce auctions, the minimum bid increment is small relative to $\pcur$ (e.g., \$0.50 for items under \$25, \$1.00 for \$25--\$100), so $\delta/\pcur$ is small by design.
Some terms are $O(\delta)$ corrections that can be omitted to first order.
Finally, we omit several mechanics such as buyer offers on zero-bid auction items, and Second Chance Offers, which are relatively uncommon and are downstream events so omitting them has negligible impact to ranking quality for this use case.

Even after these simplifications, the meCPM still requires components such as the valid-bid-through-rate $\Prob(\mathrm{valid\ bid}\mid\mathrm{imp},s)$
and the conditional bid-value CDF $F_s$.
The bid-through-rate can be fit to data analogous to standard CTR or CVR modeling, with a binary label indicating whether the impression led to a valid bid being placed. Conditional on a valid bid, the bid-value CDF $F_s$ can be estimated from the historical submitted bids in comparable auction states. This aligns with the production data-generating process: the platform can see the values of successfully submitted bids, while the bid-through-rate model captures the probability that an impression produces such a bid.
A parametric approach could model $f_s$ using a positive, right-skewed distribution such as a log-normal, gamma, Weibull, or mixture distribution, with parameters conditioned on item and auction state, and use tractable expressions for $S_s$. The integral \(\mathcal{J}_s\) could be evaluated in several ways including, for \(x \le y\), via the standard stop-loss identity
\(\mathcal{J}_s(x,y)=\E_{V\sim F_s}[(V-x)_+]-\E_{V\sim F_s}[(V-y)_+]\)
(see \cite{kaas2008modern}).
A non-parametric implementation could use the empirical plug-in estimator
\(\widehat{\mathcal{J}}_s(x,y)=\frac{1}{n}\sum_{i=1}^n\max\{0,\min(V_i,y)-x\}\)
operating directly on the discrete pool of valid bid samples
\(V_1,\ldots,V_n\) from comparable auction states, without requiring a fitted continuous distribution.
Hybrid and semi-parametric approaches are possible too, e.g.\ quantile regression to predict a small set of conditional quantiles and reconstruct $S_s$ by interpolation, or $k$-nearest-neighbor estimators that pool over the most similar historical auction states.
In all cases, individual auctions may have only a few discrete submitted bids over their lifetime, so estimation requires pooling across many items and other state features like price and time buckets. The granularity of $s$ trades off bias against variance.
Alternatively, one could bypass the full conditional bid-value distribution and summarize the price upside distribution with a point estimate, e.g.\ using $\phat(s)$, the expected final selling price~\cite{ghani2005price,vanheijst2008support} given the current state, and modeling the remaining price uplift as $\max(\phat(s) - \pcur,0)$; or conservatively assuming that any new bid only increases the price by the minimum bid increment, and using $\mathrm{meCPM}_{\mathrm{\delta}} = r \cdot \Prob(\mathrm{valid\ bid}\mid\mathrm{imp},s) \cdot \delta$.

However, these estimators cannot be fit in the cold-start setting. Instead, one possibility is to leverage an existing conversion model
but use a hybrid price variant which still captures auction dynamics while also allowing a unified treatment of the new auction and ABIN listing formats alongside fixed-price within a single meCPM score component:
\begin{equation}
\mathrm{meCPM}_{\mathrm{simplified}} \;=\; r \;\cdot p(\text{sale}\mid\text{imp}) \;\cdot \pcur
\label{eq:deployed}
\end{equation}
Comparing \eqref{eq:deployed} to the full meCPM \eqref{eq:pure_auction_exact}, this approximation involves two key substitutions.
\textbf{(i)}~$p(\text{sale}\mid\text{imp})$ replaces $p(\mathrm{valid\ bid}\mid\mathrm{imp},s)$: the existing conversion model already captures auction-relevant signals such as item quality, relevance, time dynamics, and engagement activity. Though $p(\text{sale})$ and $p(\mathrm{valid\ bid})$ predict different events, they are closely related, since high quality, relevant items likely to sell are also likely to attract bids (see Section~\ref{sec:model_eval}).
\textbf{(ii)}~For fixed-price items, $\pcur = \pFP$ and (\ref{eq:deployed}) recovers (\ref{equation:fp_ranking_score}) exactly, and for auction and ABIN items, $\pcur$ serves as the price signal instead of the full expression.
The resulting score preserves the meCPM's structural form---probability of a revenue event, times a price signal, times the ad rate---while reusing an existing well-calibrated model. Importantly, the approximation is tightest in the regime that covers the most impressions: auctions typically run for 3--10 days and bidding often concentrates near close, so a large fraction of auction impressions are in the zero-bid starting state where $\pcur = p_0$ and the theoretical meCPM reduces to $r \cdot \Prob(\mathrm{valid\ bid}\mid\mathrm{imp},s) \cdot p_0$ (Appendix~\ref{app:zero_bid}).
In this regime, the simplified form \eqref{eq:deployed}
has exactly the same structure, differing only in the substitution of $\Prob(\text{sale})$ for $\Prob(\mathrm{valid\ bid})$.
A partial additional justification is that the minimum bid increment $\delta$ is itself a monotonically increasing step function of $\pcur$ (e.g., \$0.50 at \$5--\$25, \$1.00 at \$25--\$100, \$2.50 at \$100--\$250, and so on), so the floor on incremental revenue per bid also scales with $\pcur$ (Pearson $r = 0.93$, Spearman $\rho = 0.93$, Kendall $\tau = 0.81$ between $\delta$ and $\pcur$ for prices up to \$10,000). This high rank-order correlation means using $\pcur$ in $\mathrm{meCPM}_{\mathrm{simplified}}$ roughly preserves the relative ordering of items by their minimum guaranteed uplift, similar to $\mathrm{meCPM}_{\mathrm{\delta}}$.

Finally, despite the simplicity of the form in \eqref{eq:deployed}, it serves primarily as a \emph{bootstrapping} role: it generates on-policy, auction-specific data, enabling dedicated valid-bid-through-rate models $p(\mathrm{valid\ bid}\mid\mathrm{imp},s)$, estimation of the conditional bid-value distribution $F_s$ or survival function $S_s$, or final price predictor model $\phat$, as needed for the full meCPM and its variants. As shown in Section~\ref{sec:online_results}, even this simplification delivers statistically significant improvements on live traffic.

\section{Experiments}
\subsection{Online Tests}
\label{sec:online_results}
We evaluate the proposed system through experiments on live, session randomized traffic. As seen in Table~\ref{tab:ab_results}, introducing auction listings into the ad ranking system delivered positive ad revenue gains and statistically significant ($p<.05$) improvements in engagement metrics, driven primarily by newly monetized auction-dominated queries that previously had little CPA coverage. Meanwhile, guardrail metrics were neutral or directionally positive, confirming no degradation in buyer experience.
For confidentiality, we anonymize exact metric names but note that metrics related to Search Ad Revenue, Ad CTR, Ad Relevance, and transaction volume all increased.

\begin{table}
  \caption{Online test results. Lift is relative to production control, with statistically significant lifts ($p<.05$) in bold.}
  \label{tab:ab_results}
  \centering
  \begin{tabular}{lc}
    \toprule
    Metric & Lift (\%) \\
    \midrule
    Ad Revenue Metric 1  & +0.43\% \\
    Ad Revenue Metric 2  & +1.24\% \\
    Platform Metric 1    & +1.02\% \\
    Ad Quality Metric 1  & \textbf{+0.21\%} \\
    Ad Quality Metric 2  & \textbf{+0.13\%} \\
    \bottomrule
  \end{tabular}
\end{table}

\subsection{Model Evaluation}
\label{sec:model_eval}

We evaluate model performance metrics on $>30$M logged production impressions of auction and ABIN listings at rank 1.
Discriminative performance is notable: the repurposed $p(\text{sale})$ model achieves an AUC-ROC on auction and ABIN listings that is 93\% as high as its AUC-ROC on the existing fixed-price inventory, despite never having been trained specifically for auctions, indicating reasonable transfer of the sale prediction task from existing inventory to auction items.
This is not entirely surprising since many of the signals that predict whether a fixed-price item will sell are also informative for predicting auction outcomes, such as buyer intent and item relevance and quality. Nonetheless, it demonstrates the effectiveness of this approach in addressing the cold-start problem to launch a new ad program with no prior data for these item listing types.

The discussion of calibration is more nuanced. We compute $p(\text{sale})$ calibration metrics using 10 bins of predicted probabilities and observe that MAPE on auction and ABIN listings is notably higher than on the existing fixed-price inventory. The model tended to over-predict $\Prob(\text{sale}\mid\text{imp})$ for auction listings, which is expected: per-impression conversion probability is inherently lower for auctions than for fixed-price items since they can have multiple impressions which collect bids but do not lead to an immediate purchase from that impression which removes it from the platform's inventory.
However, as in (\ref{eq:deployed}), in the component for expected revenue estimation, $\Prob(\text{sale}\mid\text{imp})$ is multiplied by other factors to get the $\mathrm{meCPM}_{\mathrm{simplified}}$, and for auction and ABIN items, since the item's current bid can only increase in time and not decrease, the overall set of impressions of an ad auction item will necessarily have a current price term which is at or below the final selling price, i.e.\ the price component of the simplified meCPM score will systematically under-predict ground-truth transaction revenue. These two points, and the fact that $p(\text{sale})$ models often have decreasing partial dependence on price, mean that various miscalibrating phenomena have opposing effects which partially offset each other.
This led to well-calibrated revenue prediction
which closely tracked the line $y=x$.
Further, since the miscalibration was very nearly linear but with reduced slope, some of the miscalibration is absorbed by other system parameters which scale the meCPM score component and combine it with other terms, and hence also provide their own implicit form of calibration,
and dilute the impact of miscalibration within the overall ranking process.
Finally, the revenue calibration is further reinforced by the sizeable share of auction impressions in the zero-bid starting state, where the simplified formula closely matches the theoretical meCPM (Section~\ref{sec:simplification}).

\section{Discussion}

Overall, our approach demonstrates that auction listings can be successfully integrated into expected value ranking with positive business and user experience impact. We highlight several considerations that are not currently limiting, but may become important as the system scales.

\paragraph{Ad System Tradeoffs}
A natural question is impression redistribution: do sponsored auction items shift away impressions from fixed-price sponsored listings? Ultimately, this comes down to net platform value: if integrating a new ad format into the ranking system delivers overall positive user engagement and platform metrics, a small shift of impression share across ad programs may be a reasonable tradeoff as long as constraints on guardrail metrics are satisfied and as long as the net effects are positive, as observed in online tests (Section~\ref{sec:online_results}).

Segmenting queries by their auction-vs-fixed-price impression mix, we observe three groups: auction-heavy queries (a small share of Search Results Pages, SRP), mixed queries (a somewhat larger share), and fixed-price-heavy queries (the large majority of SRP). Revenue gains were concentrated in the mixed-query group, while the auction-heavy and fixed-price-heavy groups showed little net change. In aggregate across all queries, the net effect on fixed-price CPA was neutral, and auction listings exhibited higher click-through rates than fixed-price listings in comparable slots, suggesting that the newly monetized inventory is largely additive rather than just a reallocation of existing demand.

\paragraph{Toward the Full Marginal eCPM Formulation}
The system described in Section~\ref{sec:simplification} is a practical and highly simplified approximation of the marginal eCPM objective derived in Section~\ref{sec:pure_auction}. A natural direction for future work is closing the gap between the two. The full formulation requires a valid-bid-through-rate estimator $\Prob(\mathrm{valid\ bid}\mid\mathrm{imp},s)$ and a conditional bid-value CDF $F_s(v) = \Prob(V \le v \mid \mathrm{valid\ bid}, s)$, which were not initially available due to the cold-start setting. As auction-specific data accumulates over time, however, the full meCPM-based approach and its variants become increasingly feasible to explore in future work, as does quantifying whether the resulting gain over the simplified score justifies the added modeling complexity.

\section{Conclusion}
We introduced marginal expected revenue per impression (meCPM) as a meaningful objective for ranking mixed listing formats within Cost-per-Acquisition e-commerce search advertising programs. Fixed-price, pure auction, and hybrid (ABIN) listings can all be unified under this generalized formulation of expected revenue in a way that naturally fits ad programs in sponsored search. Although ranking generally includes other terms and blended objectives, not solely revenue, the component for ad monetization for all listing formats can be understood as instances of a unified meCPM formulation: the probability of a revenue generating event, times the expected price uplift above locked-in revenue, times the ad rate. CPC and fixed-price CPA formats are inherently marginal, and the auction meCPM extends this to incrementally evolving prices.

The problem addressed here fits a unique space among major e-commerce platforms, as few marketplaces operate mixed fixed-price and auction listing formats at comparable scale, and even fewer integrate these formats into a unified ad ranking system.
To address the cold-start problem of launching a new ad program without prior data, we adopt a simplified approximation of this framework.
Live tests on production traffic showed statistically significant gains in user and business metrics. Future work could pursue dedicated auction ranking models for an exact marginal eCPM-based ranker, or intermediate options such as using a point estimate of final selling price to model expected remaining price uplift. Overall, our framework shows that mixed fixed-price and auction listings can be successfully integrated into expected value ranking with positive business and user experience impact.

\begin{acknowledgments}
  We acknowledge the cross-team collaboration which made this effort possible, and we thank the passionate users of the eBay platform who make it a unique marketplace for buyers and sellers.
\end{acknowledgments}

\section*{Declaration on Generative AI}
During the preparation of this work, the authors used ChatGPT 5.4/5.5 and Claude Opus 4.6/4.7 to check grammar and to suggest refinements for some sentences. After using these tools, the authors reviewed and edited the content as needed, and take full responsibility for the content in this work.

\bibliography{sample-ceur}

\appendix

\section{Notation}
\label{app:notation}

\begin{center}
\begin{tabular}{@{}c p{0.85\linewidth}@{}}
\toprule
Symbol & Meaning \\
\midrule
$r$ & Ad rate (fraction of final selling price paid by seller to platform, $r \in [0,1]$) \\
$p_{\mathrm{FP}}$ & Price of a fixed-price (FP) listing \\
$p_0$ & Starting bid price set by the seller on an auction listing \\
$\pcur$ & Current visible price on an auction listing \\
$\pR$ & Reserve price (if any) on an auction listing \\
$\pBIN$ & Buy It Now price (for ABIN listings) \\
$\delta$ & Minimum bid increment, a piecewise-constant function of price (e.g.\ \$0.50 at \$5--\$25, \$1.00 at \$25--\$100). Treated as a single value at the current $\pcur$ for the meCPM derivation; this is exact within a tier, and an approximation at tier boundaries  \\
$M$ & Current leader's hidden submitted maximum bid (undefined when no bids have been placed yet) \\
$V$ & A new potential bidder's submitted maximum bid \\
$s$ & Observable auction/context state: $\pcur$, $\delta$, bid count, time remaining, bidder-visible flags (reserve presence, reserve-met, BIN availability), and item/query/context features. The reserve price $\pR$, BIN price $\pBIN$, and hidden leader maximum $M$ are pulled out as explicit formula arguments \\
$f_s(v)$ & Conditional density of submitted maximum bids given a valid bid in state $s$ \\
$F_s(v)$ & Conditional CDF: $F_s(v) = \Prob(V \le v \mid \mathrm{valid\ bid}, s)$ \\
$S_s(v)$ & Conditional survival function: $S_s(v) = 1 - F_s(v) = \Prob(V > v \mid \mathrm{valid\ bid}, s)$ \\
$\tau$ & Valid-bid threshold: $\tau = p_0$ before any bid; $\tau = \pcur + \delta$ once auction has an active bid \\
$b$ & Lower limit of the new-leader integral: $b \;\equiv\; \max(\tau, M)$ \\
$L$ & Reserve-clearing bid threshold: $L \equiv \max\{\pR, \tau\}$ \\
$C$ & Settlement cap in reserve-not-met regime: $C \equiv \max\{\pR, M{+}\delta\}$ post-bid; $C \equiv \pR$ zero-bid \\
$\mathcal{J}_s(x,y)$ & Conditional clipped survival integral: $\mathcal{J}_s(x,y) \equiv \int_x^y S_s(v)\,dv$ if $x < y$; $\;0$ if $x \geq y$ \\
$\hat{p}_f(s)$ & Predicted (expected) final selling price for items of this type, in this state \\
\bottomrule
\end{tabular}
\end{center}

\bigskip
\section{meCPM Derivation: Auction Without Reserve Price}
\label{app:pure_auction_derivation}

We derive the marginal eCPM for an auction with no reserve price from Section~\ref{sec:pure_auction}.
Assume there is no reserve price in effect ($\pR = 0$, or equivalently $\pcur \geq \pR$).
If at least one bid has been placed, the item will sell, so the current locked-in revenue is $\pcur \cdot r$ and it will be realized regardless of whether any additional impressions are shown.

\subsection{Effect of One New Bid on the Final Price}
\label{app:pure_auction_derivation_price_change_cases}
Let $s$ denote the observable auction/context state and let $\tau$ denote the valid-bid threshold; in the post-bid setting considered here, $\tau = \pcur + \delta$. The bid-through-rate model estimates $\Prob(\mathrm{valid\ bid}\mid\mathrm{imp},s)$. Conditional on a valid bid occurring, the bidder submits a maximum $V$ drawn from the conditional CDF $F_s$, where $F_s(v) = \Prob(V \le v \mid \mathrm{valid\ bid}, s)$.
For completeness, a submitted maximum below $\tau$ would not constitute a valid bid and would not produce any price change. Since $F_s$ is conditional on a valid bid, the expectation is taken only over $V\ge \tau$, for which the visible price changes as
\begin{equation}
\Delta \pcur(V; \pcur, M, \delta)
\;=\;
\left\{
\begin{array}{lll}
(V + \delta) - \pcur & \text{if } \pcur + \delta \leq V \leq M - \delta & (\pcur \to V + \delta) \\[4pt]
M - \pcur & \text{if } M - \delta < V \leq M & (\pcur \to M) \\[4pt]
V - \pcur & \text{if } M < V < M + \delta & (\pcur \to V) \\[4pt]
(M + \delta) - \pcur & \text{if } V \geq M + \delta & (\pcur \to M + \delta)
\end{array}
\right.
\label{eq:delta_pcur_piecewise_appendix}
\end{equation}

Each case of \eqref{eq:delta_pcur_piecewise_appendix} is worth understanding:
\begin{enumerate}
\item The current leader's proxy has room to outbid the challenger: proxy bidding automatically raises the visible price from $\pcur$ to $V + \delta$. The leader remains the same.
\item The challenger's maximum is within one $\delta$ increment of $M$, so the leader's proxy wants to bid $V + \delta$ but cannot exceed its own maximum $M$. Visible price rises to $M$; leader stays the same.
\item The challenger outbids the leader ($V > M$) and takes the lead, but $V < M + \delta$, so the second-price settlement $M + \delta$ exceeds the new leader's own maximum. The visible price is capped at $V$.
\item The challenger outbids the leader by at least one full increment ($V \geq M + \delta$). The visible price rises to the old leader's maximum plus one increment, $M + \delta$.
\end{enumerate}

The four case piecewise format above is intuitive, but for convenience in later steps, we rewrite it below in a different format split into two cases by whether or not the new bid $V$ overtakes the current winning bid $M$. We write the new visible price as:

\begin{equation}
p_{\mathrm{new}}(V) =
\left\{
\begin{array}{lll}
\min(M,\, V + \delta), & \tau \leq V \leq M, & \text{(old leader)} \\[4pt]
\min(V,\, M + \delta), & V > M, & \text{(new leader)}
\end{array}
\right.
\label{eq:pnew}
\end{equation}

In the first case the current leader's proxy bids $V + \delta$ to stay ahead, capped at the leader's own maximum $M$. In the second the challenger becomes the new leader at one increment above $M$, capped at the challenger's own maximum $V$.

The price uplift is $\Delta\pcur(V) = p_{\mathrm{new}}(V) - \pcur$ for
valid bids and zero otherwise.

\subsection{Expected Incremental Revenue}
\label{app:expected_incremental_revenue}
The marginal eCPM in factored form is:
\begin{equation}
\text{meCPM}_{\text{auction}}
\;=\;
r \cdot \Prob(\mathrm{valid\ bid}\mid\mathrm{imp},s) \cdot
\E_{V\sim F_s}\!\left[\Delta \pcur(V;\pcur,M,\delta)\right]
\label{eq:pure_auction_factored_appendix}
\end{equation}
For notational simplicity, we now assume a continuous $F_s$; discrete bid grids can be handled by the corresponding sum form.
Since $F_s$ is the conditional distribution of submitted maxima given a valid bid, and valid bids satisfy $V \geq \tau$, the conditional expectation is:

\begin{equation}
\E_{V\sim F_s}[\Delta\pcur(V)]
= \int_\tau^{\infty} \Delta\pcur(V)\, f_s(V)\, dV
\label{eq:mecpm}
\end{equation}
where $f_s$ is the conditional density of submitted maximum bids given a valid bid in state $s$.

We now derive \eqref{eq:mecpm} in closed form by integrating by parts over
each branch of \eqref{eq:pnew}.
To handle regions that may be empty (when
$M$ is close to $\pcur$), we use a conditional clipped survival integral
\begin{equation}
\mathcal{J}_s(x,y)
\;\equiv\;
\begin{cases}
\displaystyle\int_x^y S_s(v)\,dv, & x < y, \\[6pt]
0, & x \geq y,
\end{cases}
\label{eq:J}
\end{equation}
which is zero whenever the integration interval is empty.

Using $\Delta\pcur(V) = p_{\mathrm{new}}(V) - \pcur$, split the integral \eqref{eq:mecpm} into the two cases of \eqref{eq:pnew}:
\[
\int_\tau^\infty \Delta\pcur(V)\,f_s(V)\,dV
\;=\;
\underbrace{%
  I_1\vphantom{\int}
}_{\text{old-leader}}
\;+\;
\underbrace{%
  I_2\vphantom{\int}
}_{\text{new-leader}}
\]
where $I_1 = \int_\tau^M (\min(M, V{+}\delta) - \pcur)\,f_s(V)\,dV$ when $\tau < M$ (and $I_1 = 0$ when $\tau \geq M$), and $I_2 = \int_{\max(\tau,M)}^\infty (\min(V, M{+}\delta) - \pcur)\,f_s(V)\,dV$.

\paragraph{Old-leader region ($I_1$, present when $\tau < M$).}

When $\tau < M$, integrate by parts with
\[
u = \min(M,\,V{+}\delta) - \pcur, \qquad dv = f_s(V)\,dV, \qquad v = -S_s(V)\!
\]
\begin{equation}
I_1
= \Bigl[-\bigl(\min(M,V{+}\delta)-\pcur\bigr)\,S_s(V)\Bigr]_\tau^M
  + \int_\tau^M \frac{d}{dV}\bigl[\min(M,\,V{+}\delta)\bigr]\;S_s(V)\,dV
\label{eq:ibp1}
\end{equation}
The left term of \eqref{eq:ibp1} evaluates to
\[
-\underbrace{\bigl(\min(M,M{+}\delta)-\pcur\bigr)}_{=\,M-\pcur}\,S_s(M)
\;+\;
\underbrace{\bigl(\min(M,\tau{+}\delta)-\pcur\bigr)}_{=\;p_{\mathrm{new}}(\tau)\,-\,\pcur}\,S_s(\tau)
\]
Under the continuous-$F_s$ convention, there is no point mass at the valid-bid threshold, so $P(V=\tau \mid \mathrm{valid\ bid}, s)=0$. Since $F_s$ is conditional on valid bids $(V \geq \tau)$, this gives $F_s(\tau)=0$ and hence $S_s(\tau)=1-F_s(\tau)=1$, so the second term simplifies to $p_{\mathrm{new}}(\tau)-p_{\mathrm{cur}}$.
For the right term of \eqref{eq:ibp1}, the derivative of $\min(M,\,V{+}\delta)$ with respect to $V$ is $1$ when
$V + \delta < M$ (i.e.\ $V < M - \delta$) and $0$ when $V + \delta > M$ (i.e.\ $V > M - \delta$).
Therefore
\begin{equation}
I_1
= \bigl(p_{\mathrm{new}}(\tau) - \pcur\bigr)
  - (M - \pcur)\,S_s(M)
  + \mathcal{J}_s(\tau,\,M{-}\delta)
\label{eq:I1}
\end{equation}

\paragraph{New-leader region ($I_2$).}

Use the same integration by parts, now with
$u = \min(V,\,M{+}\delta) - \pcur$,
$dv = f_s(V)\,dV$,
$v = -S_s(V)$,
and writing $b \equiv \max(\tau, M)$ for the lower limit:
\begin{equation}
I_2
= \Bigl[-\bigl(\min(V,M{+}\delta)-\pcur\bigr)\,S_s(V)\Bigr]_b^\infty
  + \int_b^\infty \frac{d}{dV}\bigl[\min(V,\,M{+}\delta)\bigr]\;S_s(V)\,dV
\label{eq:ibp2}
\end{equation}
For the first term of \eqref{eq:ibp2}, at $V \to \infty$ the boundary term vanishes (bounded integrand times
$S_s(V) \to 0$)
and at $V = b$,
$\min(b,\,M{+}\delta) = b$, since
$b = \max(\tau, M) \leq M + \delta$ (because $\tau = \pcur + \delta \leq M + \delta$).
So the left term contributes $(b - \pcur)S_s(b)$.
For the second term of \eqref{eq:ibp2}, the derivative of $\min(V,\,M{+}\delta)$ is $1$ when $V < M + \delta$
and $0$ when $V > M + \delta$. Therefore
\begin{equation}
I_2
= (b - \pcur)\,S_s(b)
  + \mathcal{J}_s(b,\,M{+}\delta)
\label{eq:I2}
\end{equation}

\paragraph{Combining $I_1$ and $I_2$.}

Add \eqref{eq:I1} and \eqref{eq:I2}. When $\tau < M$, $b = M$ and the
terms $-(M{-}\pcur)\,S_s(M)$ from $I_1$ and $+(b{-}\pcur)\,S_s(b)$ from
$I_2$ cancel, giving:
\begin{equation}
I_1 + I_2
\;=\;
\bigl(p_{\mathrm{new}}(\tau) - \pcur\bigr)
\;+\;
\mathcal{J}_s(\tau,\,M{-}\delta)
\;+\;
\mathcal{J}_s(M,\,M{+}\delta)
\label{eq:combined_taultM}
\end{equation}
When $\tau \geq M$, $I_1 = 0$ (the integration interval $[\tau, M]$ is empty, per its piecewise definition above) and $I_2$ alone gives the result (with $b = \tau$, so $p_{\mathrm{new}}(\tau) - \pcur = b - \pcur$):
\begin{align}
I_1 + I_2
&= 0 + (\tau - \pcur)\,S_s(\tau) + \mathcal{J}_s(\tau,\,M{+}\delta) \nonumber \\
&= (\tau - \pcur) + \mathcal{J}_s(\tau,\,M{+}\delta) \nonumber \\
&= \bigl(p_{\mathrm{new}}(\tau) - \pcur\bigr) + \mathcal{J}_s(\tau,\,M{+}\delta)
\label{eq:combined_taugeqM}
\end{align}
where we use $S_s(\tau) = 1$ (since $F_s$ is assumed continuous and is conditional on a valid bid, i.e.\ $V \geq \tau$), and $p_{\mathrm{new}}(\tau) = \tau$ from the new-leader case of \eqref{eq:pnew} (since $\tau \geq M$, and $\tau = \pcur + \delta \leq M + \delta$ gives $\min(\tau, M{+}\delta) = \tau$).
So, in both cases \eqref{eq:combined_taultM} and \eqref{eq:combined_taugeqM}, the sum of $I_1 + I_2$ is:
\begin{equation}
I_1 + I_2
\;=\;
\bigl(p_{\mathrm{new}}(\tau) - \pcur\bigr)
\;+\;
\mathcal{J}_s(\tau,\,M{-}\delta)
\;+\;
\mathcal{J}_s\bigl(\max(\tau,M),\,M{+}\delta\bigr)
\label{eq:combined}
\end{equation}
where the 2nd term vanishes when $\tau \geq M$ (interval $[\tau, M{-}\delta]$ is empty, by definition of \eqref{eq:J}), and the 3rd term unifies both cases via $b = \max(\tau, M)$.
Therefore:
\begin{equation}
\boxed{\;\;
\begin{aligned}
\mathrm{meCPM}_{\mathrm{auction}}
\;=\; & r \cdot \Prob(\mathrm{valid\ bid}\mid\mathrm{imp},s) \cdot \Bigl[ \bigl(p_{\mathrm{new}}(\tau) - \pcur\bigr) \\
& \quad +\; \mathcal{J}_s(\tau,\,M{-}\delta) \;+\; \mathcal{J}_s\bigl(\max(\tau,M),\,M{+}\delta\bigr) \Bigr]
\end{aligned}
\;\;}
\label{eq:result}
\end{equation}
in the post-bid setting (where $\tau = \pcur + \delta$).

\subsection{Sanity Checks}
\label{section:mecpm_auction_derivation_sanity_checks}

\textbf{Non-negativity.} Each component of \eqref{eq:result} is non-negative, as we would expect from an equation for marginal revenue: $\Prob(\cdot), r \in [0,1]$; $p_{\mathrm{new}}(\tau) \geq \pcur$ by definition; and $\mathcal{J}_s(x,y) \geq 0$. Therefore meCPM $\geq 0$, with units of price per impression (multiply by 1000 for Cost-per-Mille).

\textbf{Monotonicity in $\pcur$.} For a \emph{fixed} conditional distribution $F_s$, raising $\pcur$ increases $\tau = \pcur + \delta$, which shrinks both $\mathcal{J}_s$ intervals from the left, and the baseline term $(p_{\mathrm{new}}(\tau) - \pcur)$ also weakly decreases.
However, the conditional distribution $F_s$ itself depends on the auction state $s$, which includes $\pcur$: raising $\pcur$ raises the valid-bid threshold, changing the distribution that $S_s$ and $\mathcal{J}_s$ are computed from. Therefore the conditional expected uplift (the bracket term in \eqref{eq:result}) is not guaranteed to be monotonically decreasing in $\pcur$. The full meCPM also depends on how $\Prob(\mathrm{valid\ bid}\mid\mathrm{imp},s)$ varies with the auction state, so strict monotonicity of the meCPM in $\pcur$ requires conditions on both $F_s$ and $\Prob(\mathrm{valid\ bid}\mid\mathrm{imp},s)$.
Qualitatively, items whose visible price has already climbed close to their expected final price may tend to have lower marginal value per impression, and we might expect the bid-through-rate to have a decreasing partial dependence on $\pcur$ in many regimes. However, via possible \emph{social-proof} effects, where prospective bidders interpret a higher $\pcur$ or a larger accumulated bid count as a signal of item desirability and become more likely to bid themselves, may flip the dependence to increasing in some regions. In such cases, the full meCPM might not decrease monotonically in $\pcur$.

\textbf{Monotonicity in $M$.} Because $M$ is hidden from bidders and is not included in the observable state $s$, the conditional bid-value distribution $F_s$ and the valid-bid-through-rate $\Prob(\mathrm{valid\ bid}\mid\mathrm{imp},s)$ do not depend on $M$. Thus the monotonicity enters only through the integral bounds of the price-uplift term. In the bracket of \eqref{eq:result}, raising $M$ extends the upper limit of $\mathcal{J}_s(\tau, M{-}\delta)$, adding $S_s(M{-}\delta)\,dM$ of area, while shifting $\mathcal{J}_s(\max(\tau,M), M{+}\delta)$ rightward, which changes it by $(S_s(M{+}\delta) - S_s(M))\,dM$. The net rate of change is $S_s(M{-}\delta) - S_s(M) + S_s(M{+}\delta) \geq 0$, since $S_s$ is decreasing and non-negative. The term $(p_{\mathrm{new}}(\tau) - \pcur)$ is also non-decreasing in $M$ through $p_{\mathrm{new}}(\tau)$.
So overall, meCPM is monotonically increasing in $M$, which makes sense because impressions of listings with more hidden upside $M$ have more marginal value to the platform.

\textbf{Term magnitudes.} The first and third bracket terms of \eqref{eq:result} are each $O(\delta)$, while the second term $\mathcal{J}_s(\tau, M{-}\delta)$ scales with $M - \pcur$ and dominates when $M$ is significantly above $\pcur$.

\textbf{Extreme case $M = \pcur$.} Both $\mathcal{J}_s$ terms vanish and $p_{\mathrm{new}}(\tau) - \pcur = \delta$, so any valid bid would raise the visible price by exactly $\delta$: $\mathrm{meCPM} = r \cdot \Prob(\mathrm{valid\ bid}\mid\mathrm{imp},s) \cdot \delta$.

\subsection{Zero-Bid Starting Case}
\label{app:zero_bid}
The derivation of Appendix~\ref{app:pure_auction_derivation_price_change_cases} and Appendix~\ref{app:expected_incremental_revenue} above assumes at least one bid has been placed (so $M$ is defined, with $M \geq \pcur$). In the zero-bid starting case, no leader exists and any valid bid $V \geq p_0$ becomes the first leader at a visible price of $p_0$. The marginal revenue is the full settlement price (since locked-in revenue is zero), so $\text{meCPM}_\text{zero-bid} = r \cdot \Prob(\mathrm{valid\ bid}\mid\mathrm{imp},s) \cdot p_0$.

\bigskip
\section{meCPM Derivation: Auction With Reserve Price}
\label{app:reserve_derivation}

We extend the pure-auction derivation of Appendix~\ref{app:pure_auction_derivation} to auctions with a reserve price
$\pR$.
The item sells only if the winning bid exceeds $\pR$.
This creates the two regimes below.

\subsection{Reserve Already Met (\texorpdfstring{$\pcur \geq \pR$}{p\_cur >= p\_R})}

If the reserve is already met, i.e.\ the current visible price $\pcur$ is greater than the reserve price $\pR$, the item will sell and locked-in revenue is
$\pcur \cdot r$. The situation is identical to the pure-auction case of \eqref{eq:result}.

\subsection{Reserve Not Yet Met (\texorpdfstring{$\pcur < \pR$}{p\_cur < p\_R})}
If the reserve is not met, locked-in revenue is \emph{zero} since if the auction ended now there would be no sale. We start with the post-bid sub-case here, where $M$ is defined and $M < \pR$. The zero-bid sub-case is recovered from the same formula. A new bid with $V < \pR$
does not clear the reserve, so under the greedy next-impression only assumption (i) in Section~\ref{sec:pure_auction}, the incremental revenue is zero. (We note that alternative modeling approaches, e.g.\ a dynamic-programming formulation, could assign non-zero value to such bids, since raising $\pcur$ toward $\pR$, even without clearing it, brings the auction closer to a state where a future bid does clear the reserve and unlocks revenue.)
On the other hand, a new bid with $V \geq \pR$
clears the reserve and takes the lead (since $V \geq \pR > M$ in this post-bid sub-case). The
second-highest maximum is $M < \pR$, so the settlement price under the proxy-bidding rule (second case of \eqref{eq:pnew}) would be $\min(V, M{+}\delta)$, but now in the reserve price setting is floored at $\pR$, i.e.\ the platform sets $\pcur$ to the larger of the proxy-bidding price and $\pR$, so:

\[
p_{\mathrm{settle}}(V) \;=\; \max\!\bigl(\pR,\;\min(V, M{+}\delta)\bigr)
\]

Since locked-in revenue is zero in the reserve-not-met regime, under the greedy next-impression only assumption (i) in Section~\ref{sec:pure_auction}, the marginal revenue from a reserve-clearing bid is the full settlement price $p_{\mathrm{settle}}(V) \cdot r$. The meCPM is therefore $r \cdot \Prob(\mathrm{valid\ bid}\mid\mathrm{imp},s)$ times the expected settlement price over valid bids, accounting for the reserve-clearing condition. To compute this expectation, we first simplify $p_{\mathrm{settle}}(V)$.

Define $C \equiv \max\{\pR, M{+}\delta\}$. For reserve-clearing bids ($V \geq \pR$), this simplifies to $p_{\mathrm{settle}}(V) = \min(V, C)$: if $M + \delta \geq \pR$ then $C = M + \delta$ and the reserve floor has no effect, so $p_{\mathrm{settle}} = \min(V, M{+}\delta) = \min(V, C)$; if $M + \delta < \pR$ then $C = \pR$ and $p_{\mathrm{settle}} = \max(\pR, \min(V, M{+}\delta)) = \pR = \min(V, C)$ since $V \geq \pR = C$. So we can write $p_{\mathrm{settle}} = \min(V, C)$.

Next, let $\tau$ be the valid-bid threshold: $\tau = p_0$ if no valid bid has been placed yet, and $\tau = \pcur + \delta$ once the auction has an active bid. Conditional on a valid bid, $V \sim F_s$. Only valid bids with $V \geq L \equiv \max\{\pR, \tau\}$ are both valid and reserve clearing. The expected settlement price contribution over valid bids is:
\[
\E_{V \sim F_s}\bigl[\min(V, C) \cdot \ind\{V \geq L\}\bigr]
\;=\;
\int_L^\infty \min(V, C)\,f_s(V)\,dV
\]
To evaluate this integral, split at $C$:
\begin{equation}
\int_L^\infty \min(V, C)\,f_s(V)\,dV
\;=\;
\int_L^C V\,f_s(V)\,dV
\;+\;
C\int_C^\infty f_s(V)\,dV
\label{eq:reserve_split}
\end{equation}
where the first integral on the right-hand side of \eqref{eq:reserve_split} is present only when $L < C$ (and is zero when $L \geq C$).
For the first integral, integrate by parts with $u = V$, $du = dV$, $dv = f_s(V)\,dV$, $v = -S_s(V)$:
\[
\int_L^C V\,f_s(V)\,dV
\;=\;
\bigl[-V\,S_s(V)\bigr]_L^C + \int_L^C S_s(V)\,dV
\;=\;
-C\,S_s(C) + L\,S_s(L) + \mathcal{J}_s(L,\,C)
\]
where $\mathcal{J}_s$ is the conditional clipped survival integral defined in \eqref{eq:J}.
By the definition of survival function, the second integral of \eqref{eq:reserve_split} evaluates to $C\,S_s(C)$.
So, adding the two terms of \eqref{eq:reserve_split}, the $C\,S_s(C)$ terms cancel, and we get:
\[
\int_L^\infty \min(V, C)\,f_s(V)\,dV
\;=\;
L\,S_s(L) + \mathcal{J}_s(L,\,C)
\]
The above assumes $L < C$. When $L \geq C$, every reserve-clearing bidder has $V \geq L \geq C$, so $\min(V,C) = C$ throughout and the integral evaluates to $C\,S_s(L)$. Both cases ($L < C$ and $L \geq C$) are captured by the expression $\min(L,C)\,S_s(L) + \mathcal{J}_s(L,C)$, since when $L \geq C$ we have $\min(L,C) = C$ and $\mathcal{J}_s(L,C) = 0$.
So, the meCPM in the reserve-not-met regime is $r$ times $\Prob(\mathrm{valid\ bid}\mid\mathrm{imp},s)$ times this expected settlement contribution:
\begin{equation}
\boxed{\;\;
\text{meCPM}_{\text{reserve not met}}
\;=\;
r \cdot \Prob(\mathrm{valid\ bid}\mid\mathrm{imp},s) \cdot
\left[\min(L,C)\,S_s(L) \;+\; \mathcal{J}_s(L,\,C)\right]
\;\;}
\label{eq:app_reserve_not_met}
\end{equation}
where $L \equiv \max\{\pR,\tau\}$ and $C \equiv \max\{\pR, M{+}\delta\}$.
Intuitively, the first term $\min(L, C)\,S_s(L)$ is the floor settlement value (either $L$ itself or the cap $C$, whichever is smaller) earned conditional on the bid clearing $L$, and $\mathcal{J}_s(L, C)$ is the additional expected upside from $V$ exceeding $L$ on its way up to $C$.

The derivation above assumes $M$ is defined (i.e.\ at least one bid has been placed). In the zero-bid sub-case ($M$ undefined, no incumbent), the piecewise definition gives $C \equiv \pR$; combined with $L = \pR$ (since $\tau = p_0 < \pR$ in this regime), the formula collapses to $r \cdot \Prob(\mathrm{valid\ bid}\mid\mathrm{imp},s) \cdot \pR \cdot S_s(\pR)$, the marginal value of a reserve-clearing first bid, which pays exactly $\pR$ in the absence of an existing bid.

\subsubsection{Common Case}
\label{app:reserve_common_case}
Post-bid, we are in the reserve-not-met regime with $\pR > M$. In all but the narrow edge case where $M$ is within one bid increment of $\pR$ (i.e.\ $M < \pR < M + \delta$), we have $\pR \geq M + \delta$ and thus $C = \pR \leq L$, so $\mathcal{J}_s(L,C) = 0$ and the formula \eqref{eq:app_reserve_not_met} reduces to $r \cdot \Prob(\mathrm{valid\ bid}\mid\mathrm{imp},s) \cdot \pR \cdot S_s(L)$ --- the reserve price weighted by the probability that a valid bid also clears the reserve. In the narrow edge case $M + \delta > \pR$, $\mathcal{J}_s(L, C)$ adds a small additional upside (bounded by $\delta$).
Thus, the marginal eCPM in the reserve-not-met regime is driven mainly by the value of just clearing the reserve at $\pR$ (the first term of \eqref{eq:app_reserve_not_met}), with the second term adding at most a $\delta$-sized correction in the narrow edge case where $M$ is within one increment of $\pR$.

\bigskip
\section{Summary}

The following table collects the marginal eCPM formulas for all listing types.

\bigskip

\begin{center}
\begin{tabular}{lp{9cm}}
\toprule
\textbf{Listing Type} & \textbf{Marginal eCPM} \\
\midrule
Fixed Price &
$r \cdot \Prob(\text{sale}\mid\text{imp}) \cdot \pFP$ \\

Auction, post-bid, no reserve &
{\raggedright $r \cdot \Prob(\mathrm{valid\ bid}\mid\mathrm{imp},s) \cdot \bigl[\bigl(p_{\mathrm{new}}(\tau) - \pcur\bigr) + \mathcal{J}_s(\tau,\,M{-}\delta) + \mathcal{J}_s(\max(\tau,M),\,M{+}\delta)\bigr]$\par} \\

Auction, zero-bid, no reserve &
$r \cdot \Prob(\mathrm{valid\ bid}\mid\mathrm{imp},s) \cdot p_0$ \\

Auction, post-bid, reserve already met &
Same as Auction, post-bid, no reserve \\

Auction, post-bid, reserve not met &
$r \cdot \Prob(\mathrm{valid\ bid}\mid\mathrm{imp},s) \cdot \bigl[\min(L,C)\,S_s(L) + \mathcal{J}_s(L,C)\bigr]$ \\

Auction, zero-bid, reserve not met &
$r \cdot \Prob(\mathrm{valid\ bid}\mid\mathrm{imp},s) \cdot \pR \cdot S_s(\pR)$ \\

ABIN &
$r \cdot \Prob(\text{BIN}\mid\text{imp},s) \cdot \pBIN \;+\; \mathrm{meCPM}_{\mathrm{auction}}(s)$,

where $\mathrm{meCPM}_{\mathrm{auction}}(s)$ is whichever auction-format row above applies to the listing's current state \\
\bottomrule
\end{tabular}
\end{center}

\end{document}